\documentstyle[12pt,amsbsy]{article}

\def\bra#1{\left\langle #1\right|}
\def\ket#1{\left| #1\right\rangle}
\def\n#1{{\hbox{\it \bf#1}}}
\def\beq{\begin{equation}}
\def\eeq{\end{equation}}
\def\bsigma{\boldsymbol\sigma}
\def\btau{\boldsymbol\tau}
\def\bphi{\boldsymbol\phi}

\newcommand{\ci}[1]{\cite{#1}}
\newcommand{\bi}[1]{\bibitem{#1}} 
\begin{document}
 
\setcounter{page}{1}
 
\title{ Meson Exchange Currents in Pion Double Charge Exchange at High Energies}
 
\author{L. Alvarez-Ruso and M.J. Vicente Vacas
\\
{\small Departamento de F\'{\i}sica Te\'{o}rica and IFIC,
      Centro mixto Universidad de Valencia-CSIC,}
\\
{\small 46100, Burjassot, Spain}}
\maketitle
 
\vspace{1cm}
 
\begin{abstract}
 In this letter we study the high energy behavior of the forward differential
cross section for the $^{18}O(\pi^+,\pi^-)^{18}Ne$ double charge exchange 
reaction. We have evaluated the sequential and the meson exchange current
mechanisms. The meson exchange current 
contribution shows a very weak energy dependence and becomes relevant at 
incident pion kinetic energies above 600 MeV.

\vspace{0.5cm}

\end{abstract}
 
\vspace{1cm}

\newpage

Pion double charge exchange  (DCX) has generated a significant amount
of theoretical and experimental work in the last years \cite{rev1,rev2}.
The main reason is its special sensitivity to the two-nucleons
wave function because, in contrast to most nuclear reactions, which are 
dominated by one-nucleon mechanisms, DCX needs at least two nucleons to take 
place.

Up to now, most of the research has concentrated in the 
region of energies around the $\Delta$ resonance or below it. 
At these energies, the analysis of the reaction is very complicated. 
This is due to the strong distortion of the pion waves at 
resonance and the diversity of mechanisms that play a significant role :
successive deltas \cite{dint1,dint2}, meson exchange currents 
(MEC)\cite{ger,mik1,kol1,vicente},
absorption mechanisms \cite{kol2,absorption}, and many others.
Normally, the reaction is dominated by the sequential mechanism (SEQ), in
which the incoming pion undergoes two sequential single
charge exchange (SCX) scatterings. 

A recent DCX calculation, which included only the
SEQ mechanism, has been performed for incident pion energies up to  
1.4 GeV\ci{oshi}. It is shown there that the  angular cross section
at forward angles decreases above 600 MeV, reaching a first dip
at 700 MeV and a second, very pronounced, at 1300 MeV.
These very low values of the cross section, produced by the mechanism
which dominates the reaction at resonance, open the possibility for 
alternative mechanisms to show up. The clear discrepancies observed 
between the experimental results on inclusive DCX at high energies
and the predictions of a sequential charge exchange model\ci{abram}
tend to confirm these expectations.

  In this paper, we report a new DCX calculation including meson
exchange currents (fig. 1b,c) in addition to the SEQ mechanism (fig. 1a).
MEC in DCX reactions were first studied by Germond, Robilotta and Wilkin 
\ci{ger}. They found a small contribution to the cross section 
in the $\Delta$ resonance region. 
Many other calculations \cite{mik1,kol1,vicente} have followed their 
pioneering work, studying also the low energy region, and finding,
in all cases, that MEC processes were small compared to the SEQ
mechanism. However, the amplitude of the MEC diagrams considered in 
this paper depends very weakly on the energy of the incoming pion. 
Thus, one can expect these mechanisms to become important at high 
energies, in the regions where the SEQ process presents a dip.

Our approach is essentially the same as in ref.\ci{oshi}. 
We calculate the cross section for the $^{18}O(\pi^+,\pi^-)^{18}Ne$ 
reaction using the Glauber model of multiple scattering.
Assuming that the DCX process takes place  in the valence neutrons, 
whereas the core ($^{16}O$) is only responsible for the distortion of 
the pion waves, the amplitude for the DCX reaction can be written as
\beq
F({\n {q}})=\frac{ik}{2\pi }\int d^2{b}\,e^{i{\n {qb}}}\,
\Gamma_{DCX}({\n {b}})\,\Gamma_{DIS}({\n {b}}), 
%\end{equation}
\eeq
where ${\n {b}}$ is the impact parameter, $k$ the incident pion  
momentum, and ${\n {q}}$ the momentum transfer. Both  $k$  and  
${\n {q}}$   are in the laboratory system. In Eq. (1), 
$\Gamma_{DIS}$  is the distorted profile given by
\beq
\Gamma_{DIS}({\n {b}})={\bra {^{16}O}}\prod_{i\in core}(1-\Gamma^{s}({\n {b}}-
{\n {s}}_{i} )){\ket {^{16}O}},
\end{equation}
where $\Gamma^{s}$ is the isoscalar profile function. 
The quality of the approximation implicit in eq. (1), where 
the profile is factorized into a DCX and a distortion piece was discussed 
in  refs. \ci{dint2,vicente}.

We evaluate  $\Gamma_{DIS}$ following the method explained in ref.\ci{fran}.
We obtain 
\beq
\Gamma_{DIS}=(1-\Gamma)^{16},
\end{equation}
with
\beq
\Gamma=\frac{1}{2\pi ik_{cm}}\int d{^2 {q}}\,e^{-i{\n {qb}}}\,S(q)f^{s}(q),
\end{equation}
where $k_{cm}$ is the pion momentum  in the  $\pi N$ c.m. system,
$f^{s}(q)$ is the non-spin-flip isoscalar scattering amplitude (also in the
$\pi N$ c.m. system), and $S(q)$ the $^{16}O$ nuclear form factor, 
obtained by the Fourier transform of the nuclear density. 

The DCX profile function ($\Gamma_{DCX}$) will be the sum of 
$\Gamma_{SEQ}$  and $\Gamma_{MEC}$, the profiles  for the sequential
and the meson exchange currents mechanisms, respectively.
For the SEQ mechanism (fig. 1a) we find 

\beq
\Gamma_{SEQ}=-\sum_{m_{1},m_{2}}\frac{(-1)^{5-m_{1}-m_{2}}}{3}
{\bra {1d_{5/2}m_{1}}}\Gamma^{v}{\ket {1d_{5/2}m_{2}}}
{\bra {1d_{5/2}-m_{1}}}\Gamma^{v}{\ket {1d_{5/2}-m_{2}}},
\end{equation}
where  we use harmonic oscillator wave functions with a parameter 
$\alpha^2=0.32$fm$^{-2}$, and

\beq
{\bra {nljm}}\Gamma^{v}{\ket {n'l'j'm'}}=\frac{1}{2\pi ik}\int d{^2 {q}}\,
e^{-i{\n {qb}}}f^{v}({\n {q}}){\bra {nljm}}e^{i{\n {qs}}}{\ket {n'l'j'm'}}.
\end{equation}
Here, $f^v(q)$ is the isovector part of the $\pi N$ amplitude, obtained
using the SAID code \ci{arndt}. We have  included partial waves up to 
$l=6$.

The MEC profile is given by 

\beq
\Gamma_{MEC}({\n {b}})=\frac{1}{2\pi ik}\int d{^2 {q}}\,e^{-i{\n {qb}}}
F({\n {q}}),
\end{equation}
where $F({\n {q}})$ is calculated in the framework of chiral perturbation
theory (ChPT) at tree level with baryons (nucleons) incorporated\ci{pich}.
In the present work we restrict ourselves to the leading order in ChPT but,
at energies above the $\Delta$ resonance, 
higher orders could be also important, as well as the explicit
inclusion of resonances such as the $\rho ( 770 )$.

For this process, the set of relevant lagrangians is the following:
\begin{eqnarray}
{\cal L}_{NN \pi}&=&-\frac{f}{\mu}\bar{\psi}\gamma^{\mu}\gamma_{5}
\btau \psi \cdot (\partial_{\mu} \bphi),\nonumber\\
{\cal L}_{NN \pi \pi \pi}&=&\frac{1}{6f^{2}_{\pi}}\frac{f}{\mu} 
(\bar{\psi} \gamma^{\mu} \gamma_{5} \btau \psi) \cdot
 [(\partial_{\mu} \bphi) \bphi^{2}-\bphi 
(\partial_{\mu} \bphi \cdot \bphi)], \\
{\cal L}_{\pi \pi \pi \pi}&=&\frac{1}{6f^{2}_{\pi}}[
(\partial_{\mu} \bphi \cdot \bphi)^{2}-\bphi^{2} 
(\partial_{\mu} \bphi)^{2}
+\frac{1}{4}
\mu^{2} \bphi^{4}]. \nonumber
\end{eqnarray}
Here, $f_{\pi}$ is the pion decay constant ($f_{\pi}=92.4 MeV$),
$\mu$ is the pion mass and $f=1.02$. From these lagrangians we obtain two
terms, the so-called pion pole term ( fig 1b ) and contact term ( fig 1c ).
The amplitude corresponding to the contact term is give by the formula:

\begin{eqnarray}
F_{CT} & = & 
\frac{1}{3}
\frac{1}{(2\pi )^{4}}\frac{1}{f_{\pi}^{2}}\left( \frac{f}{\mu}\right)^{2}
\int d^3 {p}\sum_{m_{1},m_{2}}\frac{(-1)^{5-m_{1}-m_{2}}}{6}
{\bra {1d_{5/2}m_{1}}}\bsigma (2{\n {p}}+{\n {q}})\, 
e^{i({\n {p}}-{\n {q}}){\n {x}}}
{\ket {1d_{5/2}m_{2}}}
\nonumber\\
  &  &
{\bra {1d_{5/2}-m_{1}}}{\n {$\bsigma$p}} e^{-i{\n {py}}}{\ket {1d_{5/2}-m_{2}}} 
\frac{1}{{\n {p}}^{2}+{\mu}^{2}}  \left( \frac{\Lambda^{2}-{\mu}^{2}}
{\Lambda^{2}+{\n {p}}^{2}}\right) ^{2},
\end{eqnarray}
and  the amplitude of the pole term is

\begin{eqnarray}
F_{PT}&=& \frac{2}{3}
\frac{1}{(2\pi )^{4}}\frac{1}{f_{\pi}^{2}}\left( \frac{f}{\mu}\right)
^{2} \int d{^3 {p}}\sum_{m_{1},m_{2}}\frac{(-1)^{5-m_{1}-m_{2}}}{6}
{\bra {1d_{5/2}m_{1}}}{\n {$\bsigma$p}} e^{-i{\n {px}}}{\ket {1d_{5/2}m_{2}}}
\nonumber \\
 & &
{\bra {1d_{5/2}-m_{1}}}\bsigma ({\n {p}}-{\n {q}})
e^{i({\n {p}}-{\n {q}}){\n {y}}}{\ket {1d_{5/2}-m_{2}}}
\nonumber\\
   &  &
\frac{(2\mu^{2}+{\n {pq}}-{\n {p}^{2})}}
{({\n {p}}^{2}+{\mu}^{2})(({\n {p}}-{\n {q}})^{2}+{\mu}^{2})}
\left( \frac{\Lambda^{2}-{\mu}^{2}}{\Lambda^{2}+{\n {p}}^{2}}\right)
\left(\frac{\Lambda^{2}-{\mu}^{2}}{\Lambda^{2}+({\n {p}}-{\n {q}})^{2}}\right).
\end{eqnarray}
For the cutoff parameter we take $\Lambda=1.3 GeV$. It should be always taken
into account that the contact term and pole term contributions might differ 
from each other for different representations of the pion field leading to  
different sets of lagrangians\ci{kol1}. The only magnitude that has physical
meaning is the sum of these two amplitudes. 

   The dependence of the angular cross section at zero degrees
on the incoming pion kinetic energy is presented in fig. 2.  
Our calculation for the SEQ mechanism (dashed line) uses a new improved set of 
phase shifts. As a consequence, the results differ in shape, though not 
in order of magnitude, from those of \ci{oshi} around 1300 MeV.  
Using the same phase shifts as in  \ci{oshi} we reproduce their results 
from  400 MeV  to 1400 MeV.
The effects of the renormalization of the pion isovector interaction
discussed in \ci{oshi} are not included in our calculation. That 
renormalization reduces the SEQ cross section and would make
MEC even more visible. The SEQ cross section shows a rapid decrease
starting at 600 MeV, a first dip at 700 MeV and a second and more 
pronounced dip at 1100 MeV. The cross section corresponding to
the MEC mechanisms is quite flat as a function of energy. This
reflects the very weak energy dependence of the amplitudes. In
spite of that, the MEC amplitude is too small to produce a very significant 
enhancement in the cross section. In the region of the first dip,
the cross section obtained including MEC and SEQ mechanisms
is bigger than the SEQ one by a factor of 2. The
inclusive experiments at these energies\ci{abram} show a similar, but 
more pronounced, effect. At 1030 and 1200 MeV
both mechanisms are comparable in magnitude. In the first case, the 
interference causes a strong cancelation, while in the second it is  
additive. It is important to remark that the strong energy dependence
produced by such interference effects would 
be modified by the inclusion of any other sizable mechanisms, as well as by 
any modification in the experimental $\pi N$ phase shifts.
 
 In summary, we have shown in this paper that, because of the small
DCX cross section produced by the sequential mechanism at high energies, 
meson exchange currents processes could show up in this reaction.
A simple model for these processes predicts important changes of the cross 
section above 600 MeV. At these energies, some additional MEC mechanisms 
involving the $\rho$ meson could be important.

\section*{Acknowledgments}
 
This work was partially supported by CICYT, contract AEN  93-1205.

\newpage

\section{Figure Captions}
\bigskip
\parindent 0cm

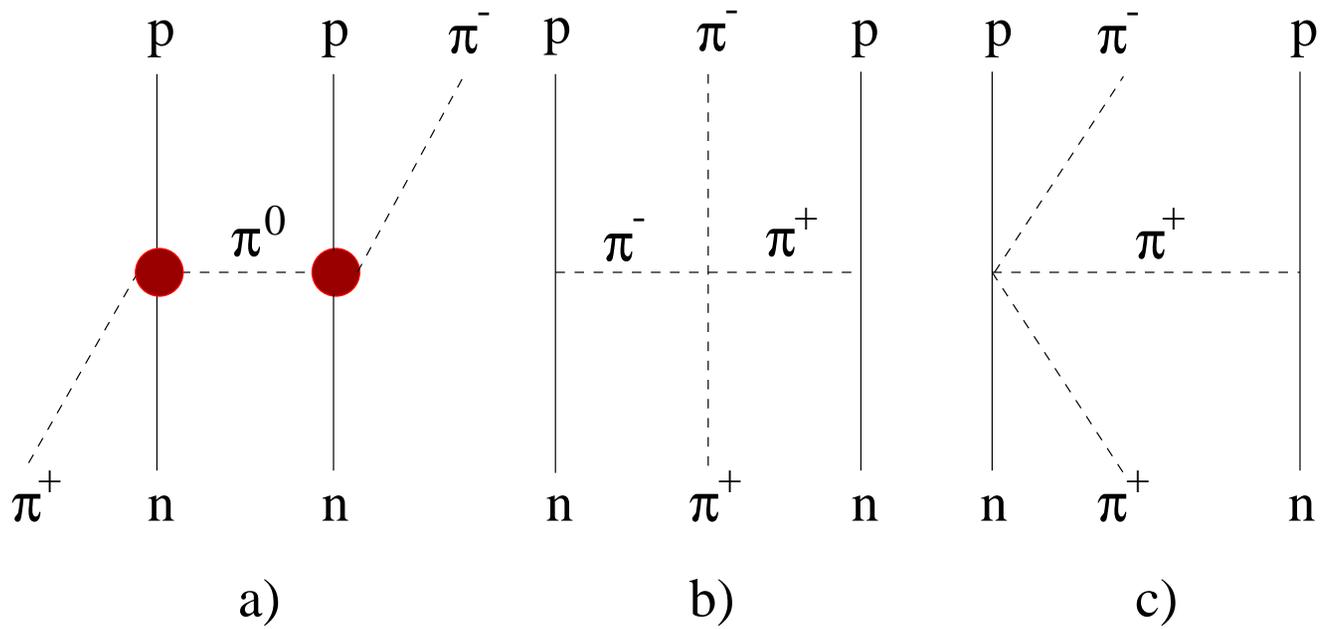
{\bf Fig. 1.} DCX mechanisms: a) Sequential, b) and c) MEC mechanisms: b) pole
term, c) contact term.
\medskip

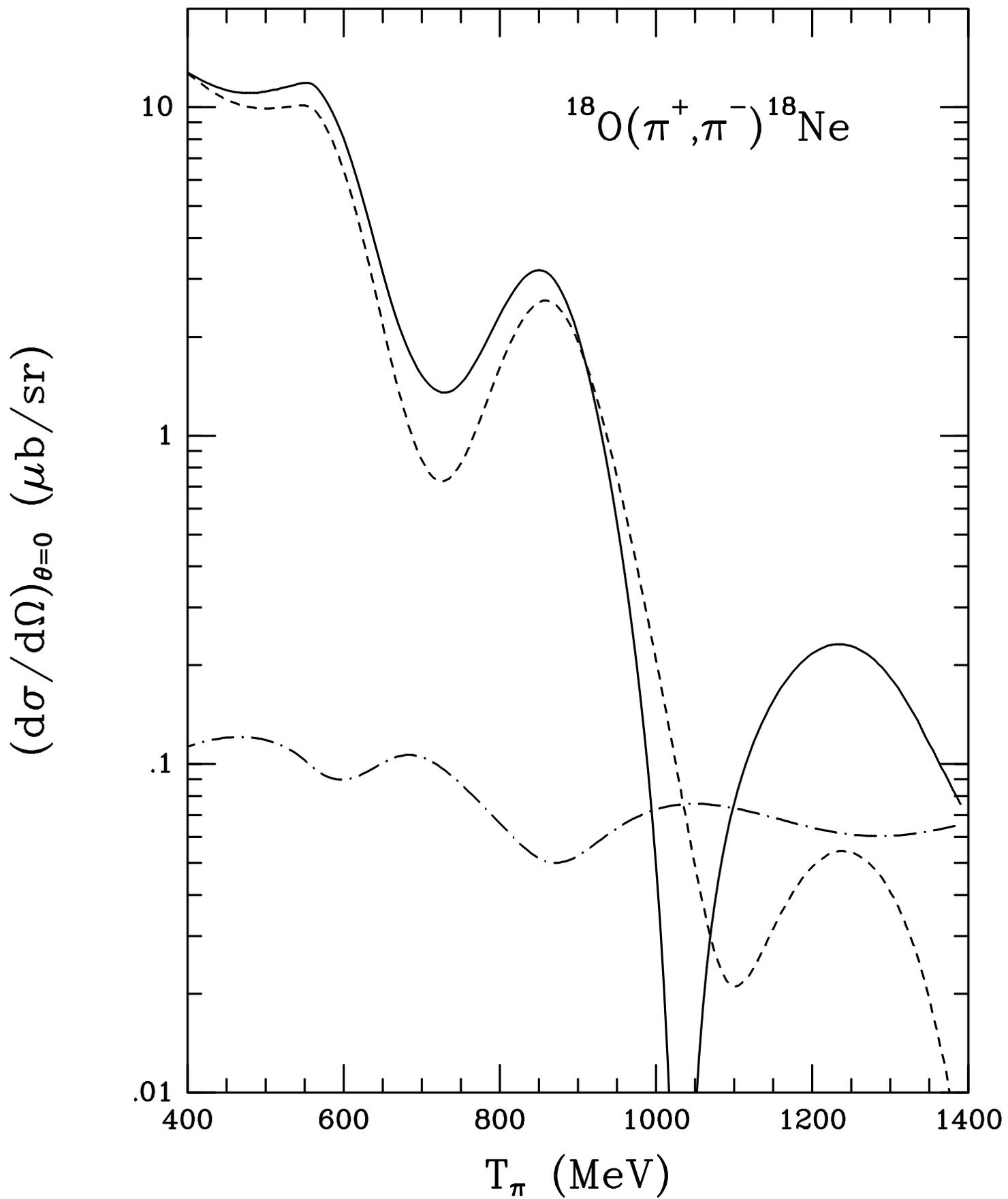
{\bf Fig. 2.} Energy dependence of the forward cross section for 
$^{18}O(\pi^+,\pi^-)^{18}Ne$. Dashed curve: SEQ mechanism, dashed-dotted line:
MEC mechanisms, full line: SEQ + MEC mechanisms.
\medskip
\end{document}